\documentclass[a4paper]{jpconf}
\usepackage{cite}
\usepackage{amsmath}%
\newtheorem{Theorem}{Theorem}
\def\spisok#1{\begin{gather}#1\end{gather}}
\def\eq#1{\begin{equation}#1\end{equation}}

\def\eqs#1{\begin{equation}\begin{split}#1\end{split}\end{equation}}
\def\seqs#1{\begin{equation*}\begin{split}#1\end{split}\end{equation*}}
\def\seq#1{\begin{equation*}#1\end{equation*}}
\def\qed{\vrule height0.6em width0.3em depth0pt}
\font\Sets=msbm10
\def\integer {\hbox{\Sets Z}}
\def\Z {\hbox{\Sets Z}}

\def\emph#1{#1}
\renewcommand{\forall}{{\hbox{ for all }}}

\begin{document}

\title{ Integrable discrete nonautonomous quad-equations as B\"acklund auto-transformations for known Volterra and Toda type semidiscrete equations}
\author{R.N. Garifullin and R.I. Yamilov}

\address{ Ufa Institute of Mathematics, Russian Academy of Sciences,\\ 112 Chernyshevsky Street, Ufa 450008, Russian Federation}
\ead{rustem@matem.anrb.ru and RvlYamilov@matem.anrb.ru}

\begin{abstract}We construct integrable discrete nonautonomous quad-equations as B\"acklund auto-transformations for known Volterra and Toda type semidiscrete equations, some of which are also nonautonomous. Additional examples of this kind are found by using transformations of discrete equations which are invertible on their solutions. In this way we obtain integrable examples of different types: discrete analogs of the sine-Gordon equation, the Liouville equation and the dressing chain of Shabat. For Liouville type equations we construct general solutions, using a specific linearization. For sine-Gordon type equations we find generalized symmetries, conservation laws and $L-A$ pairs.
\end{abstract}

\section{Introduction}

In this paper we consider discrete quad-equations 
\begin{equation}F_{n,m}(u_{n+1,m},u_{n,m},u_{n,m+1},u_{n+1,m+1})=0, \quad n,m\in \integer,\label{gF}\end{equation} which may explicitly depend on the discrete variables $n,m$. These equations are supposed to be polylinear, i.e. the functions $F_{n,m}$ are polynomials of the first order in each of their argument. Most of the known integrable equations of this kind have two generalized symmetries of the form
\spisok{\label{n-sym}\frac{du_{n,m}}{dt_1} =\phi_{n,m} (u_{n+1,m}, u_{n,m}, u_{n-1,m}),\\
\label{m-sym}\frac{du_{n,m}}{dt_2} =\psi_{n,m} (u_{n,m+1}, u_{n,m}, u_{n,m-1}),} see e.g. \cite{lpsy08,ly11,xp09,x09}. Recently a few examples of quad-equations with more complicated generalized symmetries have been found \cite{a11,gy12,mx13,shl13}, but such equations are out of consideration in this paper. 

Almost all known equations of the form \eqref{gF} possessing the symmetries (\ref{n-sym},\ref{m-sym}) are autonomous. In the essentially more difficult nonautonomous case, we study in this paper, only a few examples are known \cite{xp09}. 

The discrete equation \eqref{gF} can be interpreted as a chain of B\"acklund auto-transfor\-mations for the lattice equations (\ref{n-sym},\ref{m-sym}). Such transformations allow one to construct a new solution $u_{n,m+1}$ in the case of eq. \eqref{n-sym} and $u_{n+1,m}$ in the case of eq. \eqref{m-sym}, starting from a given solution $u_{n,m},$ see a more detailed comment in \cite{lpsy08}.

In this paper we start from known integrable equations of the form \eqref{n-sym} and look for the discrete equations \eqref{gF} generating for them chains of the B\"acklund auto-transformations. Such problem has been solved up to now only once by one example and only in the autonomous case \cite{ly09}. Discrete equations obtained in this way may be not integrable. We select integrable cases by requiring the existence of a second symmetry of the form \eqref{m-sym}. In such way we can construct integrable discrete equations, using known integrable equations of the Volterra type presented in \cite{y83,y06} or using their nonautonomous generalizations given in \cite{ly97}.

More precisely, we are going to use differential-discrete equations of the form:
\eq{\frac{d u_n}{dt}=P_n(u_n)(u_{n+1}-u_{n-1}).\label{Volt}}
It has been shown in \cite{ly97} that nonlinear integrable equations of this form are described by the following conditions: \eq{P_n=\alpha u_n^2+\beta_n u_n+\gamma_n,\label{fP}} where $\alpha$ is an arbitrary constant, and $\beta_n, \gamma_n$  are the two-periodic functions: \eq{ \beta_{n+2}=\beta_n,\quad \gamma_{n+2}=\gamma_n.\label{fP1}}
Up to the transformations $$\tilde t=\eta t,\quad \tilde{u}_n=\mu_n u_n+\nu_n,\quad \tilde u_n=u_{n+1},$$ where $\mu_{n}$ and $\nu_n$ are the two-periodic functions, we have five cases: the Volterra equation with $P_n=u_n$, its three modifications with $P_n=u_n^2$, $P_n=u_n^2-1$, $P_n=u_n^2-\chi_n$ and an equation with $P_n=\chi_n u_n+\chi_{n+1}$. The last equation is nothing but one of forms of the Toda model, as it has been shown in \cite{ly97}. Here $\chi_n$ is given by \eq{\label{def_chi}\chi_n=\frac{1+(-1)^n}2.}

We fix the generalized symmetry \eqref{n-sym} in the direction $n$ in one of the following five ways:
\spisok{\frac{du_{n,m}}{dt_1}=u_{n,m}(u_{n+1,m}-u_{n-1,m}),\label{V}\\
\frac{du_{n,m}}{dt_1}=u_{n,m}^2(u_{n+1,m}-u_{n-1,m}),\label{mV1}\\
\frac{du_{n,m}}{dt_1}=(u_{n,m}^2-1)(u_{n+1,m}-u_{n-1,m})\label{mV2},\\
\frac{du_{n,m}}{dt_1}=(u_{n,m}^2-\chi_{n})(u_{n+1,m}-u_{n-1,m})\label{mV3},\\
\frac{du_{n,m}}{dt_1}=(\chi_{n}u_{n,m}+\chi_{n+1})(u_{n+1,m}-u_{n-1,m}).\label{T}} In all these cases we find all corresponding polylinear discrete equations \eqref{gF}.
A more general form of such symmetry is possible:
\eq{\frac{du_{n,m}}{dt_1}=(\alpha_m u_{n,m}^2+\beta_{n,m} u_{n,m}+\gamma_{n,m})(u_{n+1,m}-u_{n-1,m}),\label{mVnon}}
where $\beta_{n+2,m}=\beta_{n,m}$ and $\gamma_{n+2,m}=\gamma_{n,m}$ for all $n,m.$ In this case we will construct some examples.

As a result we find nonautonomous integrable examples of several different types. According to their symmetry properties, eqs. \eqref{gF} are the discrete analogs of the hyperbolic type equations $u_{xy}=f(x,y,u,u_x,u_y).$ Some examples obtained in this paper are of the sine-Gordon type. Such equations have two generalized symmetries (\ref{n-sym},\ref{m-sym}) and are not Darboux integrable, see definitions in the next section. We also find a few Darboux integrable equations which can be called the discrete analogs of the Liouville equation. For all equations of this type, we construct general solutions. One more new interesting integrable example presented here is a discrete analog of the well-known dressing chain studied in  \cite{sy90,s92,vs93}.

In Section \ref{sec_th} we give some definitions and obtain theoretical results necessary for the paper. In Section \ref{sec_bac} we enumerate all the discrete equation \eqref{gF} corresponding to the differential-discrete eqs. (\ref{V}-\ref{T}) and also obtain some examples in the case of eq. \eqref{mVnon}.
In Section \ref{sec_lax} a discrete analog of the dressing chain is discussed. The problem of construction of the second symmetry \eqref{m-sym} for examples obtained in Section \ref{sec_bac} is solved in Section \ref{sec_sec}. In Section \ref{sec_tran} some additional examples are found by using special transformations of the discrete equations invertible on their solutions. 

\section{Theory}\label{sec_th}

In this section we give necessary definitions and derive some conditions for the discrete equations which allow us to make the class (\ref{gF}) of the discrete equations essentially more narrow. 

We consider equations of the form \eqref{gF} which are {\it polylinear} and {\it nondegenerate}. It is convenient to formulate definitions in terms of the function $$F_{n,m}(x_1,x_2,x_3,x_4)$$ which depends on 4 continuous complex variables $x_1,x_2,x_3,x_4$ and on 2 integer discrete ones $n,m$. An equation of the form \eqref{gF} is polylinear if \seq{\frac{\partial^2 F_{n,m}}{\partial{x_i^2}}=0,\ \ i=1,2,3,4,} for all $n,m\in\Z$. So, we consider a class of polynomial equations with 16  $n,m$-dependent coefficients.
The nondegeneracy is defined following \cite{ly11}. If the function $F_{n,m}$ depends on $x_4$ for all $n,m$, then we can rewrite the equation $F_{n,m}=0$ in the form \eq{x_4=f_{n,m}(x_1,x_2,x_3)\label{dis}} and we require the function $f_{n,m}$ to depend essentially on all its continuous variables  for all $n,m$. So, we have the following nondegeneracy condition in terms of $F_{n,m}$ and $f_{n,m}$: \eq{\label{nondeg}\frac{\partial F_{n,m}}{\partial{x_4}},\frac{\partial f_{n,m}}{\partial{x_1}},\frac{\partial f_{n,m}}{\partial{x_2}},\frac{\partial f_{n,m}}{\partial{x_3}}\neq0\ \ \hbox{for all }n,m\in\Z.}

The discrete equation \eqref{gF} is equivalent to
\eq{u_{n+1,m+1}=f_{n,m}(u_{n+1,m},u_{n,m},u_{n,m+1})\label{gf}.} 
The {\it compatibility conditions} of eqs. \eqref{gf} and (\ref{n-sym},\ref{m-sym}) have the form:
\spisok{\phi_{n+1,m+1}=\phi_{n+1,m}\frac{\partial f_{n,m}}{\partial u_{n+1,m}}+\phi_{n,m}\frac{\partial f_{n,m}}{\partial u_{n,m}}+\phi_{n,m+1}\frac{\partial f_{n,m}}{\partial u_{n,m+1}},\label{cond-n}\\ \psi_{n+1,m+1}=\psi_{n+1,m}\frac{\partial f_{n,m}}{\partial u_{n+1,m}}+\psi_{n,m}\frac{\partial f_{n,m}}{\partial u_{n,m}}+\psi_{n,m+1}\frac{\partial f_{n,m}}{\partial u_{n,m+1}}\label{cond-m}.} These relations are obtained by differentiating eq. \eqref{gf} with respect to the times $t_1$ and $t_2$ of eqs. (\ref{n-sym},\ref{m-sym}). For fixed values of $n,m$ we can express, using eq. \eqref{gf}, all functions $u_{n+k,m+l}$ ($k,l\neq0$) in terms of the functions \eq{u_{n+k,m},u_{n,m+l},\ \ k,l\in\integer,\label{ind_var}} which can be considered as {\it independent variables}. Eqs. (\ref{cond-n},\ref{cond-m}) must be identically satisfied for all values of the independent variables as well as for any $n,m\in\integer.$ If eqs. \eqref{gf} and (\ref{n-sym},\ref{m-sym}) are compatible, then (\ref{n-sym},\ref{m-sym}) are the generalized symmetries of \eqref{gf} and, on the other hand, eq. \eqref{gf} defines chains of the B\"acklund auto-transformations for eqs. (\ref{n-sym},\ref{m-sym}), see \cite{lpsy08}.

Eq. \eqref{gf} is called {\it Darboux integrable} if it has two first integrals $W_{n,m},V_{n,m}$ depending of a finite number of the independent variables \eqref{ind_var} and satisfying the relations
\eq{\label{darb1}(T_1-1)W_{n,m}=0,\quad (T_2-1)V_{n,m}=0,\quad \hbox{for all }n,m\in\Z,} on the solutions of eq. \eqref{gf}. Here $T_1,T_2$ are the shift operators in the first and second directions, respectively: $$T_1 h_{n,m}=h_{n+1,m},\quad T_2 h_{n,m}=h_{n,m+1}.$$ 
It is easy to show that the first integrals $W_{n,m}$ and $V_{n,m}$ depend only on the independent variables $u_{n,m+i}$ and $u_{n+j,m}$, respectively.  Applying the shift operators, we can represent these first integrals as:
\eqs{W_{n,m}=W_{n,m}(u_{n,m},u_{n,m+1},\ldots,u_{n,m+k_1}),\\
\label{darb2} V_{n,m}=V_{n,m}(u_{n,m},u_{n+1,m},\ldots,u_{n+k_2,m}).}
The Darboux integrable equations are linearizable, with linearizing transformations $w_{n,m}=W_{n,m},\ v_{n,m}=V_{n,m}$, and are analogs of the Liouville equation.

A discrete equation \eqref{gf} is of the {\it sine-Gordon type} if it has two generalized symmetries (\ref{n-sym},\ref{m-sym}) and is not Darboux integrable. Such equations should be integrable by the inverse scattering method. It is difficult to prove that a given equation has no first integrals \eqref{darb2}, see \cite{gy12u} for possible difficulties. For examples obtained below, we check that fact for $k_i\leq 4.$

Let us derive two conditions necessary for the compatibility of the discrete equation \eqref{gf} and an equation of the form \eq{\frac{du_{n,m}}{dt_1}=P_{n,m}(u_{n,m})(u_{n+1,m}-u_{n-1,m}).\label{Vgen}} The only restriction here is that $P_{n,m}(x)\neq0$ for all $n,m$. The equations (\ref{V}-\ref{mVnon}) are particular cases of eq. \eqref{Vgen}.

Differentiating the compatibility condition \eqref{cond-n} with respect to $u_{n+2,m}$ and applying $T_1^{-1}$, we obtain the relation
\eq{T_2\left(\frac{\partial \phi_{n,m}}{\partial u_{n+1,m}}\right)\frac{\partial f_{n,m}}{\partial u_{n+1,m}}=\frac{\partial \phi_{n,m}}{\partial u_{n+1,m}}T_1^{-1}\left(\frac{\partial f_{n,m}}{\partial u_{n+1,m}}\right).\label{cond_diff}}
This is nothing but one of so-called integrability conditions obtained in \cite{ly09}. Here we just present it in the most general non-autonomous case and write it down in a form more convenient for the present paper. In the case of eq. \eqref{Vgen} it takes the form:
\eq{P_{n,m+1}(u_{n,m+1})\frac{\partial f_{n,m}}{\partial u_{n+1,m}}=P_{n,m}(u_{n,m})T_1^{-1}\left(\frac{\partial f_{n,m}}{\partial u_{n+1,m}}\right).\label{cond_diff_P}}

Applying $T_1^{-1}$ to eq. \eqref{gf}, we can easily rewrite it in one more form \eq{u_{n-1,m+1}=\hat{f}_{n,m}(u_{n-1,m},u_{n,m},u_{n,m+1})\label{ump}}equivalent  to eq. \eqref{gF}. The function $\hat{f}_{n,m}$ essentially depends on all its continuous variables  for all $n,m$. The compatibility condition for \eqref{ump} and \eqref{n-sym} reads: 
\eq{\phi_{n-1,m+1}=\phi_{n-1,m}\frac{\partial \hat{f}_{n,m}}{\partial u_{n-1,m}}+\phi_{n,m}\frac{\partial \hat{f}_{n,m}}{\partial u_{n,m}}+\phi_{n,m+1}\frac{\partial \hat{f}_{n,m}}{\partial u_{n,m+1}}.\label{cond-n2}} 
The following condition analogous to \eqref{cond_diff_P} is derived from \eqref{cond-n2} in a quite similar way:
\eq{P_{n,m+1}(u_{n,m+1})\frac{\partial \hat{f}_{n,m}}{\partial u_{n-1,m}}=P_{n,m}(u_{n,m})T_1\left(\frac{\partial \hat{f}_{n,m}}{\partial u_{n-1,m}}\right).\label{cond_diff_P2}}

\begin{Theorem} If the discrete equation \eqref{gF} is compatibles with \eqref{Vgen}, then the conditions \eqref{cond_diff_P} and \eqref{cond_diff_P2} must be satisfied.\end{Theorem}

Using the conditions \eqref{cond_diff_P} and \eqref{cond_diff_P2}, we can essentially simplify the form of the polylinear discrete equation \eqref{gF}. The resulting form is
\eqs{&(\kappa_{1,n,m}u_{n,m}+\kappa_{2,n,m}u_{n,m+1}+\kappa_{3,n,m})u_{n+1,m+1}\\
  +&(\kappa_{4,n,m}u_{n,m}+\kappa_{5,n,m}u_{n,m+1}+\kappa_{6,n,m})u_{n+1,m}\\
  +&(\kappa_{7,n,m}u_{n,m}+\kappa_{8,n,m}u_{n,m+1}+\kappa_{9,n,m})=0,\label{formF}} where $\kappa_{i,n,m}$ are arbitrary $n,m$-dependent functions. This form corresponds to the following restrictions:
  \eq{\frac{\partial^2 F_{n,m}}{\partial u_{n+1,m+1}\partial u_{n+1,m}}=\frac{\partial^2 F_{n,m}}{\partial u_{n,m+1}\partial u_{n,m}}=0.}

\begin{Theorem}If a nondegenerate polylinear equation \eqref{gF} is compatible with an equation of the form \eqref{Vgen}, then it must have the form \eqref{formF}.
\end{Theorem}

\paragraph{Proof.} The relation \eqref{cond_diff_P} depends on the following independent variables: $u_{n-1,m},$ $u_{n,m},$ $u_{n,m+1},$ $u_{n+1,m},$ and only the term $\frac{\partial f_{n,m}}{\partial u_{n+1,m}}$ depends on $u_{n+1,m}$. Differentiating \eqref{cond_diff_P} with respect to $u_{n+1,m}$ and taking into account the restriction $P_{n,m}(x)\neq0,$ we obtain: \eq{\label{d2f1}\frac{\partial^2 f_{n,m}}{\partial u_{n+1,m}^2}=0.} In quite similar way the relation \eqref{cond_diff_P2} implies: \eq{\label{d2f2}\frac{\partial^2 \hat{f}_{n,m}}{\partial u_{n-1,m}^2}=0.}Applying the conditions \eqref{d2f1} and \eqref{d2f2} to the discrete polylinear equation \eqref{gF} and using its nondegeneracy, we are led to the form \eqref{formF}. \qed

\section{ B\"acklund auto-transformations}\label{sec_bac}

Here we describe all polylinear discrete equations \eqref{gF} compatible with eqs. (\ref{V}-\ref{T}). At the end we construct an example corresponding to an equation of the form \eqref{mVnon}.

We fix one of eqs. (\ref{V}-\ref{T}) as the generalized symmetry \eqref{n-sym}. To find corresponding discrete equation, we use in the first step more simple necessary conditions  (\ref{cond_diff_P},\ref{cond_diff_P2}) instead of the compatibility condition \eqref{cond-n}. The relations (\ref{cond_diff_P},\ref{cond_diff_P2}) are equivalent to a nonlinear algebraic system of equations for eighteen functions $\kappa_{i,n,m},\ \kappa_{i,n-1,m}.$ It is interesting that not only in this system but also in case of the compatibility condition \eqref{cond-n} the discrete variable $m$ is not changed. So we can define the dependence of $\kappa_{i,n,m}$ on $n$ only.

Such problem is not difficult in the autonomous case. We just need to solve an algebraic system for nine unknown coefficients $k_i$, using any computer algebra system like Reduce. In the nonautonomous case, we can only find from that algebraic system a set of solutions which have the form of relations between the functions $k_{i,n,m}$ and $k_{i,n-1,m}$ in a fixed point $n=n_0$. Another difficulty is that there are many divisors of zero $\alpha_n\not\equiv0$ and $\beta_n\not\equiv0$, such that  $\alpha_n \beta_n\equiv 0$. 

Nevertheless, comparing sets of solutions at $n=n_0$ and $n=n_0+1$, one can choose consistent pairs of nondegenerate solutions, i.e. such that corresponding discrete equation (\ref{formF}) is nondegenerate, and can determine in this way the dependence of the functions $k_{i,n,m}$ on $n$. In case of the nonautonomous equations \eqref{mV3} and \eqref{T}, we need to solve the corresponding algebraic system twice, at $n=2k_0$ and $n=2k_0+1$, to avoid a dependence on the function $\chi_n$. This the way we find the discrete equations.

In case of the Toda lattice \eqref{T}, we have checked that there is no polylinear and nondegenerate discrete equation \eqref{gF} compatible with \eqref{T}. However, in Section \ref{sec_tran} we present an example of the sine-Gordon type, corresponding to \eqref{T}, which is not polylinear. 
In case of the Volterra equation \eqref{V} we get a positive result:

\begin{Theorem}
If a polylinear nondegenerate discrete equation \eqref{gF} is compatible with eq. \eqref{V}, then, up to the  multiplication by a function $\nu_{n,m}$ nonzero for any $n,m\in\Z$, it can be expressed as:
\eqs{\Omega_{n,m}&(u_{n+1,m+1}u_{n,m+1}-u_{n+1,m}u_{n,m})\\&+\Omega_{n+1,m}(u_{n+1,m+1}+u_{n,m+1}-u_{n+1,m}-u_{n,m}+k_m)=0,\label{bacV}\\ &\qquad\qquad\qquad\Omega_{n,m}=\frac{1+\omega_m(-1)^n}2,}
where $\omega_m^2=1$ for all $m\in\Z$ and $k_m$ is an arbitrary $m$-dependent function.
\end{Theorem}

In the case of eqs. (\ref{mV1}-\ref{mV3}) the problem of finding  discrete equations \eqref{gF} is easier. We solve an algebraic system, corresponding to eqs. \eqref{cond_diff_P} and \eqref{cond_diff_P2} in a fixed point $n$, and see that all possible nondegenerate solutions have the same structure: $$k_{1,n,m}=k_{5,n,m}=0,\quad k_{4,n,m}=-k_{2,n,m}\neq0\ \  {\hbox{   for all }} n,m\in\Z.$$
Deviding the discrete equation \eqref{formF} by $k_{2,n,m}$, we obtain a very simple ansatz of the form \eqs{u_{n+1,m+1}u_{n,m+1}&-u_{n+1,m}u_{n,m}+\kappa_{3,n,m}u_{n+1,m+1}+\kappa_{6,n,m}u_{n+1,m}\\&+\kappa_{7,n,m}u_{n,m}+\kappa_{8,n,m}u_{n,m+1}+\kappa_{9,n,m}=0.\label{form_F_mV}}
Now we can interpret algebraic systems of equations for $\kappa_{i,n,m},\ \kappa_{i,n+1,m},\ \kappa_{i,n-1,m}$, corresponding to the conditions (\ref{cond_diff_P},\ref{cond_diff_P2}) and \eqref{cond-n}, 
as systems of ordinary difference equations for the functions $k_{i,n,m}$ and then we can specify eq. \eqref{form_F_mV} with no difficulties.

The resulting discrete equation in the case of \eqref{mV1} is not interesting:
\eqs{u_{n+1,m+1}u_{n,m+1}=u_{n+1,m}u_{n,m}.} It can be linearized by the point transformation $v_{n,m}=\log u_{n,m}.$

\begin{Theorem}
Up to the multiplication by a nonzero function $\nu_{n,m}$, a polylinear and nondegenerate discrete equation \eqref{gF} must be of the form:
\eqs{(u_{n+1,m}+a_{n+1,m})(u_{n,m}-a_{n,m})=(u_{n+1,m+1}+b_{n+1,m+1})(u_{n,m+1}-b_{n,m+1}),\\ a_{n+2,m}=a_{n,m},\quad b_{n+2,m}=b_{n,m},\quad a_{n,m}^2=b_{n,m}^2=1\label{bacmV2},}
if it is compatible with eq. \eqref{mV2}, and of the form: 
\eqs{(u_{n+1,m}+A_m\chi_{n+1})(u_{n,m}-A_m\chi_{n})=(u_{n+1,m+1}+B_{m+1}\chi_{n+1})(u_{n,m+1}-B_{m+1}\chi_{n}),\\ A_m^2=B_m^2=1,\label{bacmV3}} in the case of eq. \eqref{mV3}.
\end{Theorem}

Let us construct now a generalization of eqs. \eqref{bacmV2} and \eqref{bacmV3}, using a pair of Miura type transformations, see \cite{y94} and a more close to the discrete quad-equations \cite{ly09u}.

Eq. \eqref{mVnon} with $\alpha_m\neq0$ for any $m$ can be transformed into \eq{\frac{du_{n,m}}{dt_1}=(u_{n,m}^2-a^2_{n,m})(u_{n+1,m}-u_{n-1,m}), \quad a_{n+2,m}=a_{n,m}\ \forall n,m.\label{mVnona}} A transformation has the form $\tilde u_{n,m}=\mu_m u_{n,m}+\nu_{n,m}$,  where $\mu_m\neq0,\ \nu_{n+2,m}=\nu_{n,m}$ for all $n,m$. Eq. \eqref{mVnona} is transformed into the Volterra equation \eqref{V} by a Miura type transformation, such that \eq{\hat u_{n,m}= (u_{n+1,m}+a_{n+1,m})(u_{n,m}-a_{n,m}).} Introducing $b_{n,m},$ such that \seq{b_{n+2,m}=b_{n,m},\quad b_{n,m}^2=a_{n,m}^2,} we have another Miura type transformation of eq. \eqref{mVnona} into \eqref{V}:
\eq{\hat u_{n,m-1}= (u_{n+1,m}+b_{n+1,m})(u_{n,m}-b_{n,m}).} Excluding $\hat u_{n,m}$ we are led to the discrete equation \eqs{(u_{n+1,m}+a_{n+1,m})(u_{n,m}-a_{n,m})=(u_{n+1,m+1}+b_{n+1,m+1})(u_{n,m+1}-b_{n,m+1}),\\ a_{n+2,m}=a_{n,m},\quad b_{n+2,m}=b_{n,m},\quad a_{n,m}^2=b_{n,m}^2 \ \ \forall n,m.\label{bac1}} It can be checked that the quad-equation \eqref{bac1} is compatible with \eqref{mVnona}.

One can see that eqs. \eqref{bacmV2} and \eqref{bacmV3} correspond to particular cases of the general formula \eqref{bac1}. In the second case, we have $a_{n,m}^2=b_{n,m}^2=\chi_{n}$ and we can represent $$a_{n,m}=A_m\chi_n,\quad b_{n,m}=B_m\chi_n. $$

\section{Discrete analog of the dressing chain}\label{sec_lax}

Let us consider a particular case of eq. \eqref{bac1}, presented in \cite{ly09}, namely:
\eq{(u_{n+1,m}+\delta_m)(u_{n,m}-\delta_m)=(u_{n+1,m+1}-\delta_{m+1})(u_{n,m+1}+\delta_{m+1}),\label{od_d}}where $\delta_m$ is an arbitrary $m$-dependent coefficient.
It is a complete analogue of the well-known dressing chain, see \cite{sy90,s92,vs93}: \eq{\frac{d}{dx}(u_{m+1}+u_m)=u_{m+1}^2-u_m^2+\delta_{m+1}-\delta_m.\label{dch}}
Eq. \eqref{dch} can be constructed by two Miura transformations into the Korteweg-de Vries equation as well as eq. \eqref{od_d} is constructed by two discrete Miura transformations into the Volterra equation which also is called the discrete Korteweg-de Vries equation. There is in \cite{ly09} an $L-A$ pair for eq. \eqref{od_d} which is the direct analog of an $L-A$ pair for \eqref{dch} constructed in \cite{sy90}.
In this section we present a generalization of eq. \eqref{od_d} together with its $L-A$ pair.

That generalization is a particular case of eq. \eqref{bac1} corresponding to $b_{n,m}=-c_{m}a_{n,m}$ with $c_{m}^2\equiv 1$:
\eqs{&(u_{n+1,m+1}-c_{m+1}a_{n+1,m+1})(u_{n,m+1}+c_{m+1}a_{n,m+1})\\&=(u_{n+1,m}+a_{n+1,m})(u_{n,m}-a_{n,m}),\quad \label{od_d_ob} a_{n+2,m}=a_{n,m},\ \ c_{m}^2\equiv 1,} and we lose e.g. the case $b_{n,m}=(-1)^na_{n,m}$. In the case when $a_{n,m}\neq0$ and $c_m=1$ for all $n,m$, we can introduce a new function $\hat u_{n,m}$: $u_{n,m}=\hat u_{n,m} a_{n,m}/\delta_m,$ where $\delta_m^2=a_{n+1,m}a_{n,m}$. The last product does not depend on $n$, as $a_{n,m}$ is two-periodic with respect to $n$. This function $\hat u_{n,m}$ satisfies an equation of the form \eqref{od_d}, i.e. we get nothing new in this case. So, only the case when $c_m=-1$ or $a_{n,m}=0$ for some $n,m$ is interesting, and we will show examples of this kind in next sections.

An $L-A$ pair for eq. \eqref{od_d_ob} reads:
\eqs{L_{n,m}=&\left(\begin{array}{cc}\lambda& -(u_{n+1,m}+a_{n+1,m})(u_{n,m}-a_{n,m})\\ 1&0\end{array}\right),\\
A_{n,m}=&\left(\begin{array}{cc}\lambda& a_{n+1,m+1}(u_{n,m+1}+a_{n,m+1})(1+c_{m+1})\\ -\frac{a_{n,m+1}(1+c_{m+1})}{u_{n,m+1}-a_{n,m+1}}& \frac{\lambda (u_{n,m+1}+c_{m+1}a_{n,m+1})}{u_{n,m+1}-a_{n,m+1}} \end{array}\right),\label{la_od}}
and these matrices satisfy the standard relation $L_{n,m+1}A_{n,m}=A_{n+1,m}L_{n,m}.$ In partilucar case \eqref{od_d}, this $L-A$ pair coincides with one of \cite{ly09}. A hierarchy of conservation laws for eq. \eqref{od_d} has been constructed in \cite{hy13} by using that $L-A$ pair of \cite{ly09}. 

We also can construct conservation laws for \eqref{od_d_ob}, using the $L-A$ pair (\ref{la_od}) and the same scheme of \cite{hy13}.
Discrete conservation laws are of the form
\eqs{(T_2-1)p_{n,m}^{(i)}=(T_1-1)q_{n,m}^{(i)}\label{con_law}}
and, for $i=0,1,2$, are given by the following functions $p_{n,m}^{(i)}$ and $q_{n,m}^{(i)}$:
\eqs{p_{n,m}^{(0)}&=\log(u_{n+1,m}+a_{n+1,m})(u_{n,m}-a_{n,m}),\\ q_{n,m}^{(0)}&=\log\frac{u_{n,m+1}+c_{m+1}a_{n,m+1}}{u_{n,m+1}-a_{n,m+1}};}
\eqs{p_{n,m}^{(1)}&=(u_{n+1,m}+a_{n+1,m})(u_{n,m}-a_{n,m}),\\ q_{n,m}^{(1)}&=-(u_{n,m+1}-a_{n,m+1})(1+c_{m+1})a_{n+1,m+1};}
\eqs{p_{n,m}^{(2)}&=(u_{n+1,m}+a_{n+1,m})(u_{n,m}-a_{n,m})(2u_{n+2,m}u_{n+1,m}+u_{n+1,m}u_{n,m}\\&-2u_{n+2,m}a_{n+1,m}+u_{n+1,m}a_{n,m}+a_{n+1,m}u_{n,m}-3a_{n+1,m}a_{n,m}) ,\\
q_{n,m}^{(2)}&=-2(1+c_{m+1})a_{n+1,m+1}(u_{n,m+1}-c_{m+1}a_{n,m+1})(u_{n+1,m}u_{n,m}-u_{n+1,m}a_{n,m}\\&+a_{n+1,m+1}u_{n,m+1}+a_{n+1,m}u_{n,m}-a_{n+1,m+1}a_{n,m+1}-a_{n+1,m}a_{n,m})\label{pq2}.}

Equations of the sine-Gordon type possess two hierarchies of generalized symmetries and conservation laws. Eq. \eqref{od_d_ob} probably has, in general, only one hierarchy of conservation laws and generalized symmetries, see a comment below. Nevertheless, it has the $L-A$ pair and is integrable for this reason. It deserves further study as a direct discrete analogue of the dressing chain \eqref{dch}.

The discrete equation \eqref{od_d} has the generalized symmetry \eqref{mVnona} with $a_{n,m}=\delta_m$ in the direction $n$.   However, in the case when $\delta_m\neq0\ \forall m,$ we can show that if the second generalized symmetry of the form \eqref{m-sym} exists, then the following relation must take place: \seq{\delta_m^2=\delta_0^2\ \forall m.} In this case, using rescaling $u_{n,m}=\hat u_{n,m} \delta_m,$ we can transform this equation into: \eq{(u_{n+1,m+1}-1)(u_{n,m+1}+1)=(u_{n+1,m}+1)(u_{n,m}-1)\label{od_d1}.}

This is the well-known integrable equation found in \cite{ht95,nc95}. A generalized symmetry in the $m$-direction can be found in \cite{ly09}, $L-A$ pairs of different forms have been presented e.g. in \cite{nc95,ly09u}.

A new	sine-Gordon type example of the form \eqref{od_d_ob}, which is essentially nonautonomous, will be discussed in Section \ref{sec_tran}. That equation has two hierarchies of generalized symmetries as well as two hierarchies of conservation laws which can be derived from the $L-A$ pair \eqref{la_od}.

\section{Second generalized symmetry}\label{sec_sec}

Among discrete equations, we consider in this paper, there may be integrable equations which do not have two generalized symmetries of the form (\ref{n-sym},\ref{m-sym}). Nevertheless, we use here the existence of two generalized symmetries of such form as an integrability criterion. It is constructive and allows us not only to check an equation for integrability, but also to solve some classification problems, as it is demonstrated below, see also \cite{gy12}.

For some of discrete equations found in Section \ref{sec_bac}, we select in this section cases in which a generalized symmetry of the form \eqref{m-sym} exists. The symmetry must be {\it nondegenerate}, i.e. its right hand side must differ from zero for any $n,m.$ 

When searching for generalized symmetries, we use a scheme developed in \cite{ly09,ly11,ggh11}. To check the Darboux integrability of an equation, we use results of \cite{h05,gy12}. In both cases some special annihilation operators introduced in \cite{h05} play an important role.

\subsection{Volterra case}\label{Volterracase}
In the case of eq. \eqref{bacV} we have two possibilities up to the transformation $\tilde u_{n,m}=u_{n+1,m}$. 
We have $k_m\equiv 0$ in both cases. In the first case $\omega_m=(-1)^m$, therefore $\Omega_{n,m}=\chi_{n+m}$, and eq. \eqref{bacV} takes the form:
\eqs{\chi_{n+m}&(u_{n+1,m+1}u_{n,m+1}-u_{n+1,m}u_{n,m})\\&+\chi_{n+m+1}(u_{n+1,m+1}+u_{n,m+1}-u_{n+1,m}-u_{n,m})=0\label{bacV_1},} where $\chi_{n+m}$ is defined by \eqref{def_chi}. Nonautonomous integrable equations of this kind are known, see \cite{xp09}. Examples of \cite{xp09} and eq. \eqref{bacV_1} are essentially different, as corresponding generalized symmetries strongly differ from each other.

The second generalized symmetry \eqref{m-sym} in the $m$-direction of eq. \eqref{bacV_1} reads:
\eq{\frac{du_{n,m}}{dt_2}=u_{n,m}\left(\frac{C_m}{u_{n,m+1}-u_{n,m}}+\frac{C_{m-1}}{u_{n,m}-u_{n,m-1}}\right),\quad C_m=\alpha m+\beta,\label{fir_V}} where $\alpha,\ \beta$ are arbitrary constants. 
Eq. \eqref{fir_V} with $\alpha=0,\beta=1$ is a representative of the well-known complete list of integrable Volterra type equations presented in \cite{y83,y06}. 
Eq. \eqref{fir_V} with $\alpha=1,\beta=0$ is its master symmetry found in \cite{cy96}. It generates  generalized symmetries for eq. \eqref{fir_V} with $\alpha=0,\beta=1$ and, therefore, for the discrete equation \eqref{bacV_1}, for instance:
\eqs{\frac{du_{n,m}}{d\tau_2}&=\frac{u_{n,m+1}u_{n,m}}{(u_{n,m+2}-u_{n,m+1})(u_{n,m+1}-u_{n,m})^2}+\frac{u_{n,m}u_{n,m-1}}{(u_{n,m}-u_{n,m-1})^2(u_{n,m-1}-u_{n,m-2})}\\&+\frac{u_{n,m}^2(u_{n,m+1}-u_{n,m-1})}{(u_{n,m+1}-u_{n,m})^2(u_{n,m}-u_{n,m-1})^2}.}

It can be proved that eq. \eqref{bacV_1} does not have first integrals \eqref{darb2} with $k_i\leq 4$ and, for this reason, probably is not  Darboux integrable.

In the second case, $\omega_m\equiv 1$ in eq. \eqref{bacV}, hence $\Omega_{n,m}=\chi_{n}$, and this equation is of the form:
\eqs{\chi_{n}&(u_{n+1,m+1}u_{n,m+1}-u_{n+1,m}u_{n,m})\\&+\chi_{n+1}(u_{n+1,m+1}+u_{n,m+1}-u_{n+1,m}-u_{n,m})=0\label{bacV_2}.} Its generalized symmetry of the form \eqref{m-sym} reads:
\eq{\frac{du_{n,m}}{dt_2}=c_m\frac{(u_{n,m+1}-u_{n,m})(u_{n,m}-u_{n,m-1})}{u_{n,m+1}-u_{n,m-1}},\label{sec_V}}
where $c_m$ is an arbitrary function, such that $c_m\neq 0$ for any $m.$ This example is Darboux integrable, as it possesses the following first integrals, see (\ref{darb1},\ref{darb2}):
\eqs{V_{n,m}=\chi_{n}u_{n+1,m}u_{n,m}+\chi_{n+1}(u_{n+1,m}+u_{n,m}), \\ W_{n,m}=\frac{(u_{n,m+3}-u_{n,m})(u_{n,m+2}-u_{n,m+1})}{(u_{n,m+3}-u_{n,m+2})(u_{n,m+1}-u_{n,m})}.\label{first_I}} In the next section, a general solution of eq. \eqref{bacV_2} will be constructed.

There exist examples with degenerate generalized symmetries, too. Such examples can be taken from the following statement: eq. \eqref{sec_V} is the symmetry of eq. \eqref{bacV} iff $$c_m(\omega_m-\omega_{m-1})=c_m k_m=c_m k_{m-1}=0 \ \forall m.$$ There may be very few points of degeneration of a symmetry, for instance, if \seq{k_m\equiv 0, \quad \omega_m=-1,\ m\leq-1,\quad \omega_m=1,\ m\geq 0,} then $c_0=0$ and $c_m$ may be a nonzero constant in all the other points $m$. Such examples probably have not been considered in the literature up to now. Equations of this kind seem to be very close to the integrable ones and, in our opinion, deserve further study.

\subsection{General solution}\label{sec_gen}
In this section we present, by an example, a scheme of the construction of a general solution for the Darboux integrable discrete equation. We use an observation of \cite{gy12} that, in many cases, first integrals of the first order (i.e. such that $k_1=1$ or $k_2=1$ in \eqref{darb2}) can be rewritten as a linear equation. Using this fact, corresponding discrete equation can be equivalently rewritten as a nonautonomous linear equation.

We see that $V_{n,m}$ of \eqref{first_I} is a first integral of the first order, and the relation $(T_2-1)V_{n,m}=0$ is equivalent to the discrete equation \eqref{bacV_2}. We can solve this relation and obtain
\eqs{V_{n,m}=\chi_{n}u_{n+1,m}u_{n,m}+\chi_{n+1}(u_{n+1,m}+u_{n,m})=\eta_n,} where $\eta_n$ is an arbitrary $n$-dependent function of integration.

We express the function $\eta_n$ in the form 
\eqs{\eta_n = \chi_n\rho_n\rho_{n+1}+\chi_{n+1}(\rho_n+\rho_{n+1}),} with a new arbitrary function $\rho_n$, 
 and introduce $v_{n,m}$ so that
\eqs{u_{n,m} = \frac1{v_{n,m}}+\rho_{n}.\label{fromutov}}
For this new unknown function $v_{n,m}$ we obtain a linear equation:
\eqs{(\chi_{n+1}+\chi_n\rho_n)v_{n,m}+(\chi_{n+1}+\chi_n\rho_{n+1})v_{n+1,m}+\chi_n=0.\label{eq_v}}
Introducing the notation \eqs{\rho_{n}=\chi_n a_n+\chi_{n+1}b_{n},} we can define a linear transformation: \eq{\label{fromvtow}v_{n,m}=w_{n,m}+\chi_n \alpha_n-\chi_{n+1}\alpha_{n+1},} where $\alpha_n$ is defined by 
\eqs{ b_n=\frac{a_{n-1}\alpha_{n-1}+1}{\alpha_{n+1}}.}
Now the linear equation \eqref{eq_v} turns into the homogeneous linear one: \eqs{(\chi_{n+1}+a_n\chi_n)w_{n,m}+(\chi_{n+1}+b_{n+1}\chi_n)w_{n+1,m}=0.} 

By using an integrating factor $\chi_n\nu_n+\chi_{n+1}\mu_n,$ we can represent this equation as a total difference:  \eqs{(T_1-1)\Xi_{n,m}=0,\quad \Xi_{n,m}=(\kappa_{n-1}\chi_{n}-\kappa_{n}\chi_{n+1})w_{n,m}} if \eqs{\alpha_{n}=\frac{\kappa_{n}}{\kappa_{n-1}},\quad a_n=\frac{\kappa_{n-1}}{\nu_n},\quad \nu_n=\kappa_{n+2}-\kappa_{n},\quad \mu_n = -\kappa_n.}
We are led to the relation $\Xi_{n,m}=\theta_m$ with a new function of integration $\theta_m$. From this relation we find
\eqs{w_{n,m}= \frac{\theta_m}{\kappa_{n-1}\chi_{n}-\kappa_{n}\chi_{n+1}}} and, taking into account (\ref{fromutov},\ref{fromvtow}), we obtain  \eqs{u_{n,m}=\chi_n\frac{\kappa_{n-1}(\theta_m+\kappa_{n+2})}{(\kappa_{n+2}-\kappa_n)(\theta_m+\kappa_n)}+\chi_{n+1}\frac{\kappa_{n}(\theta_m+\kappa_{n-1})}{(\kappa_{n+1}-\kappa_{n-1})(\theta_m+\kappa_{n+1})}.\label{genu}}

This formula \eqref{genu} defines the general solution of eq. \eqref{bacV_2} in the sense that this solution depends on two arbitrary functions  of one discrete variable: $\kappa_n,\theta_m$.

\subsection{Modified Volterra case}\label{modi}
Here we consider the discrete equations (\ref{bacmV2},\ref{bacmV3}) corresponding to the modified Volterra equations (\ref{mV2},\ref{mV3}).

As $b_{n,m}$ of eq. \eqref{bacmV2} is a 2-periodic function with respect to $n$, we can introduce $d_{m}=b_{n,m}b_{n+1,m}$ which does not depend on $n$. After rescaling $u_{n,m}\to -u_{n,m}b_{n,m},\ a_{n,m}\to -a_{n,m}b_{n,m}$, we obtain instead of (\ref{mV2},\ref{bacmV2}) the following equations:
\eqs{\frac{du_{n,m}}{dt_1}=d_m(u_{n,m}^2-1)(u_{n+1,m}-u_{n-1,m})\label{mV2p},}
\eqs{d_{m+1}(u_{n+1,m+1}-1)(u_{n,m+1}+1)=d_m (u_{n+1,m}+a_{n+1,m})(u_{n,m}-a_{n,m})\label{dmV2},} where $d_m^2=a_{n,m}^2=1,\ a_{n+2,m}=a_{n,m}$ for all $n,m.$

There are here two cases with a nondegenerate second generalized symmetry. In both cases $d_m\equiv 1$ and $a_{n,m}=A_m,$ i.e. it does not depend on $n$. In the first case $A_m\equiv 1$, and eq. \eqref{dmV2} is nothing but the known equation \eqref{od_d1}. In the other case $A_m\equiv -1$, and the second generalized symmetry reads: \eqs{\frac{du_{n,m}}{dt_2}=c_m\frac{(u_{n,m+1}-u_{n,m})(u_{n,m}-u_{n,m-1})}{u_{n,m+1}-u_{n,m-1}},} where $c_m$ is an arbitrary function, such that $c_m\neq 0$ for any $m.$ This discrete equation coincides, up to an autonomous M\"obius transformation, with an example found in \cite{s10}. This equation is Darboux integrable, its first integrals and a general solution are presented in \cite{s10}, too.

There is a degenerate example of the form \eqref{dmV2} with $d_m\equiv 1,\  a_{n,m}=A_m$: 
\eqs{(u_{n+1,m+1}-1)(u_{n,m+1}+1)=(u_{n+1,m}+A_{m})(u_{n,m}-A_{m})\label{dmV3},}
where $A_m$ is such that $A_m^2=1$ and it is not constant. Eq. (\ref{dmV3}) has a second symmetry of the form
\eqs{\frac{du_{n,m}}{dt_2}&=c_m(A_m-1)(A_{m-1}-1)\frac{(u_{n,m+1}-u_{n,m})(u_{n,m}-u_{n,m-1})}{u_{n,m+1}-u_{n,m-1}}\\ &+\hat{c}_m(A_m-1)(A_{m-1}+1)\frac{(u_{n,m+1}-u_{n,m})(u_{n,m}+u_{n,m-1})}{u_{n,m+1}+u_{n,m-1}},} with two arbitrary functions $c_m,\hat{c}_m$, such that $c_m\neq 0$, $\hat{c}_m\neq 0$ for any $m$. If $A_m=1$ for some $m=M$, then $\frac{du_{n,M}}{dt_2}=0$. For example, if $A_0=1$ and $A_m=-1$ for $m\neq0$, then $\frac{du_{n,m}}{dt_2}=0$ iff $m=0$.

In case of the discrete equation \eqref{bacmV3}, we cannot enumerate all particular cases possessing the second symmetry of the form \eqref{m-sym}. We just discuss here two Darboux integrable examples. 

The first of them is eq. \eqref{bacmV3} with $A_m\equiv B_m\equiv -1$:
\eqs{(u_{n+1,m}-\chi_{n+1})(u_{n,m}+\chi_{n})=(u_{n+1,m+1}-\chi_{n+1})(u_{n,m+1}+\chi_{n}).\label{bacmV3_1}}
Its first integrals read:
\eqs{&V_{n,m}=(u_{n+1,m}-\chi_{n+1})(u_{n,m}+\chi_{n}), \label{W1_V31}\\&W_{n,m}=\frac{(u_{n,m+1}-u_{n,m-1})(u_{n,m+1}\chi_n+u_{n,m}\chi_{n+1})}{u_{n,m+1}(u_{n,m}-u_{n,m-1})}.}
Following the scheme described in Section \ref{sec_gen}, we construct a general solution with two arbitrary functions $\kappa_n,\theta_m$:
\eqs{u_{n,m}=\chi_n\frac{(\kappa_n+\kappa_{n+2})\theta_m-1}{(\kappa_{n+2}-\kappa_n)\theta_m}+\chi_{n+1}\frac{\kappa_n\theta_m}{2\kappa_{n+1}\theta_m-1}.\label{sol_V3_1}}
The discrete equation (\ref{bacmV3_1}) is a particular case of eq. \eqref{od_d_ob} corresponding to  $a_{n,m}=-\chi_n$ and $c_m\equiv -1$. We have $q^{(i)}_{n,m}\equiv 0$ for all three conservation laws given by (\ref{con_law}-\ref{pq2}). So, all these conservation laws turn into first integrals, providing $V_{n,m}$ of (\ref{W1_V31}), in particular. This is natural for the Darboux integrable equations.

The second case corresponds to $A_m\equiv 1,\ B_m\equiv -1$ in eq. \eqref{bacmV3}, and this equation  takes the form
\eqs{(u_{n+1,m}+\chi_{n+1})(u_{n,m}-\chi_{n})=(u_{n+1,m+1}-\chi_{n+1})(u_{n,m+1}+\chi_{n}).\label{bacmV3_2}} The first integrals in this case are more complicated:
\eqs{V_{n,m}=u_{n+1,m}u_{n-1,m}(u_{n,m}^2-1)\chi_n+(u_{n+1,m}+u_{n-1,m})u_{n,m}\chi_{n+1},\\
W_{n,m}=(u_{n,m+1}+u_{n,m})(u_{n,m}+u_{n,m-1})\left(\frac{\chi_n}{u_{n,m}^2-1}+\frac{\chi_{n+1}}{u_{n,m+1}u_{n,m-1}}\right).	\label{fimV3}}
This discrete equation is a particular case of eq. \eqref{od_d_ob} corresponding to  $a_{n,m}=\chi_n$ and $c_m\equiv 1$. It can be checked that the conservation law \eqref{pq2} becomes trivial in the sense that $p_{n,m}^{(2)}$ is expressed in terms of a first integral and a total difference:
\eq{p_{n,m}^{(2)}=T_1(2V_{n,m}\chi_{n}+V_{n,m}^2\chi_{n+1})+(1-T_1)(u_{n,m}^2(u_{n+1,m}+1)^2\chi_{n+1}),} and  conservation laws \eqref{con_law} with $i\geq 3$ are trivial in the same sense, too.
As the first integrals \eqref{fimV3} are not of the first order, we cannot apply the scheme of Section \ref{sec_gen}. A general solution for eq. \eqref{bacmV3_2} will be constructed in the next section in a different way.

\section{Transformations}\label{sec_tran}

In this section we construct a few new integrable discrete examples, using equations found in Section \ref{Volterracase} and some special transformations invertible on solutions of discrete equations. Those transformations provide Miura type transformations for both discrete equations and their generalized symmetries. We use a transformation theory developed in \cite{y90,y93,y94}. A formulation more close to the discrete equations together with some applications can be found in \cite{s10}.

\subsection{Transformation 1}\label{sec_tran1}

Let us start from eq. \eqref{bacV} with $k_m\equiv 0$. Multiplying by the function $\frac{\Omega_{n,m}}{u_{n,m+1}u_{n+1,m}}+\Omega_{n+1,m}$, which is nonzero for any $n,m$, we can rewrite the equation as:
\eqs{\Omega_{n,m}&\frac{u_{n+1,m+1}}{u_{n+1,m}}+\Omega_{n+1,m}(u_{n+1,m+1}-u_{n+1,m})\\=&\,\Omega_{n,m}\frac{u_{n,m}}{u_{n,m+1}}+\Omega_{n+1,m}(u_{n,m}-u_{n,m+1}),\label{bacVm}\\ \Omega_{n,m}&=\frac{1+\omega_m(-1)^n}2,\quad \omega_m^2\equiv 1.}
Denoting the right hand side of eq. \eqref{bacVm} by $v_{n,m}$, we get two formulas
\eqs{v_{n-1,m}&=\Omega_{n-1,m}\frac{u_{n,m+1}}{u_{n,m}}+\Omega_{n,m}(u_{n,m+1}-u_{n,m}),\\ \label{uv} v_{n,m}&=\Omega_{n,m}\frac{u_{n,m}}{u_{n,m+1}}+\Omega_{n-1,m}(u_{n,m}-u_{n,m+1})} relating the functions $u_{n,m}, u_{n,m+1}$ and $v_{n,m},v_{n-1,m}.$ These formulas define a transformation of the discrete equation \eqref{bacVm} invertible on its solutions. The inverse transformation has the form:
\eqs{\label{vu}u_{n,m}&=-\Omega_{n-1,m}\frac{v_{n,m}}{v_{n-1,m}-1}-\Omega_{n,m}\frac{v_{n,m}v_{n-1,m}}{v_{n,m}-1}, \\ u_{n,m+1}&=-\Omega_{n,m}\frac{v_{n-1,m}}{v_{n,m}-1}-\Omega_{n-1,m}\frac{v_{n,m}v_{n-1,m}}{v_{n-1,m}-1}.}
After an additional point transformation:
\eqs{w_{n,m}=2\Omega_{n,m}v_{n,m}+\Omega_{n+1,m}\frac{1+v_{n,m}}{1-v_{n,m}},\label{vw}} we obtain the following discrete equation:
\eqs{(w_{n+1,m+1}-\Omega_{n+1,m+1})(w_{n,m+1}+\Omega_{n,m+1})=(w_{n,m}-\Omega_{n,m})(w_{n+1,m}+\Omega_{n+1,m}).\label{dw}}
The generalized symmetry \eqref{V} of eq. \eqref{bacVm} turns into
\eq{\frac{dw_{n,m}}{dt_1}=(w_{n,m}^2-\Omega_{n,m})(w_{n+1,m}-w_{n-1,m})\label{nw}} up to scaling $t_1$, and this is a symmetry of eq. \eqref{dw} in the $n$-direction. 

It should be remarked that the symmetry \eqref{nw} is a particular case of eqs. (\ref{mVnon},\ref{mVnona}). The discrete equation \eqref{dw} is a particular case of eq. (\ref{bac1}) and of eq. \eqref{od_d_ob} with $c_m\equiv 1$ and has, for this reason, an $L-A$ pair of the form \eqref{la_od}.

As it has been stated in Section \ref{Volterracase}, there are two subcases among eqs. (\ref{bacV})  possessing the second generalized symmetry of the form \eqref{m-sym}.
The first of them is given by $k_m\equiv 0$, $\Omega_{n,m}=\chi_{n}$ and is eq. (\ref{bacV_2}). In this case eq. \eqref{dw} coincides with eq. \eqref{bacmV3_2} and is Darboux integrable. Using the composition of transformations (\ref{uv},\ref{vw}) and the solution \eqref{genu} of eq. (\ref{bacV_2}), we obtain a general solution for eq. \eqref{bacmV3_2}:
\eqs{w_{n,m}=\chi_n\left(1-\frac{2(\theta_m+\kappa_{n+2})(\theta_{m+1}+\kappa_n)}{(\kappa_{n+2}-\kappa_n)(\theta_{m+1}-\theta_m)}\right)-\chi_{n+1}\frac{2\kappa_n(\theta_{m+1}-\theta_m)}{(\theta_{m+1}+\kappa_{n+1})(\theta_m+\kappa_{n+1})}.\label{sol_dw}}

The second case is given by $\Omega_{n,m}=\chi_{n+m}$, and eq. \eqref{dw} takes the form: 
\eqs{(w_{n+1,m+1}-\chi_{n+m})(w_{n,m+1}+\chi_{n+m+1})=(w_{n,m}-\chi_{n+m})(w_{n+1,m}+\chi_{n+m+1}).\label{dw1}} 
Its second symmetry corresponding to \eqref{fir_V} reads:
\eqs{\frac{dw_{n,m}}{dt_2}=&\chi_{n+m+1}(C_{m-1}w_{n,m-1}-C_{m+1}w_{n,m+1})\\-&\chi_{n+m}\frac{(C_{m-1}w_{n,m+1}-C_{m+1}w_{n,m-1})(w_{n,m}^2-1)}{w_{n,m+1}w_{n,m-1}},} where $C_m = \alpha m + \beta$.
After the point transformation $\xi_{n,m}=\chi_{n+m}w_{n,m}+\chi_{n+m+1}/w_{n,m}$, this symmetry takes the form:
\eqs{\frac{d\xi_{n,m}}{dt_2}=(\xi_{n,m}^2-\chi_{n+m})(C_{m+1}\xi_{n,m+1}-C_{m-1}\xi_{n,m-1}).\label{msymxi}} Eq. \eqref{msymxi} with $C_m\equiv 1$ and eq. \eqref{nw}  with $\Omega_{n,m}=\chi_{n+m}$ are the same up to the transformation $n\leftrightarrow m$. Eq. \eqref{msymxi}  with $C_m\equiv 1$ is, for any fixed value of $n$, a ${1+1}$-dimensional differential-difference equation equivalent to the well-known modified Volterra equation \eqref{Volt} with $P_n(u_{n})=u_n^2-\chi_n$. Eq. \eqref{msymxi}  with $C_m\equiv m$ is the master symmetry of eq. \eqref{msymxi}  with $C_m\equiv 1$. So, we can see that the discrete equation \eqref{dw1} possesses two hierarchies of generalized symmetries in both directions $n$ and $m$.

Eq. \eqref{dw1} is a particular case of \eqref{od_d_ob} with $c_m\equiv 1$ and $a_{n,m}=\chi_{n+m}$ and possesses, for this reason, an $L-A$ pair of the form \eqref{la_od}. In addition to \eqref{con_law} we can construct in this case the second hierarchy of conservation laws
\eqs{(T_1-1)\hat q_{n,m}^{(i)}=(T_2-1)\hat p_{n,m}^{(i)}\label{con_law2},} using this $L-A$ pair and the scheme of \cite{hy13}. First three conservation laws \eqref{con_law2} of eq. \eqref{dw1} are defined by:
\seqs{\hat q_{n,m}^{(0)}=&\log\left(\frac{\chi_{n+m}}{1-w_{n,m}}+\chi_{n+m+1}w_{n,m}\right),\\
\hat p_{n,m}^{(0)}=&-\chi_{n+m+1}\log(w_{n,m}(1-w_{n+1,m}));\\
\hat q_{n,m}^{(1)}=&\frac{1-w_{n,m}}{1+w_{n,m}}\chi_{n+m}-\frac{1+w_{n,m+1}}{w_{n,m}}\chi_{n+m+1},\\
\hat p_{n,m}^{(1)}=&\frac{2}{w_{n+1,m}}\chi_{n+m};\\
\hat q_{n,m}^{(2)}=&\frac{(1+w_{n,m+2})(1-2w_{n,m}-w_{n,m+2})}{w_{n,m+1}^2}\chi_{n+m}\\+&\frac{(1-w_{n,m+1})(2w_{n,m+2}w_{n,m+1}+2w_{n,m+2}+w_{n,m+1}w_{n,m}-w_{n,m})}{w_{n,m+2}^2w_{n,m}}\chi_{n+m+1},\\
\hat p_{n,m}^{(2)}=&\frac{4w_{n+1,m}(w_{n,m+1}+1)^2}{(1+w_{n+1,m})^2w_{n,m}^2}\chi_{n+m+1}.}

\subsection{Transformation 2}\label{sec_tran2}
Let us start now from eq. \eqref{bacV_1} and rewrite it in the form:
\eqs{\chi_{n+m}&{u_{n+1,m+1}}{u_{n,m+1}}+\chi_{n+m+1}(u_{n+1,m+1}+u_{n,m+1})\\=&\chi_{n+m}{u_{n+1,m}}u_{n,m}+\chi_{n+m+1}(u_{n+1,m}+u_{n,m}).\label{bacVmTT}}
Denoting the right hand side of \eqref{bacVmTT} by $U_{n,m+1}$, we get two formulas
\eqs{U_{n,m}=\chi_{n+m+1}u_{n+1,m}u_{n,m}+\chi_{n+m}(u_{n+1,m}+u_{n,m}),\\ U_{n,m+1}=\chi_{n+m}u_{n+1,m}u_{n,m}+\chi_{n+m+1}(u_{n+1,m}+u_{n,m}),\label{uU}}
which define a transformation invertible on the solutions of eq. \eqref{bacVmTT}. The inverse transformation is given by:
\eqs{u_{n,m}=-\frac{\chi_{n+m+1}}2\left(U_{n,m+1}+\sqrt{U_{n,m+1}^2-4U_{n,m}}\right)-\frac{\chi_{n+m}}2\left(U_{n,m}+\sqrt{U_{n,m}^2-4U_{n,m+1}}\right),\\
u_{n+1,m}=\frac{\chi_{n+m+1}}2\left(\sqrt{U_{n,m+1}^2-4U_{n,m}}-U_{n,m+1}\right)+\frac{\chi_{n+m}}2\left(\sqrt{U_{n,m}^2-4U_{n,m+1}}-U_{n,m}\right).\label{Uu}}

In terms of new function $U_{n,m}$, we obtain the following discrete equation:
\eqs{&\chi_{n+m+1}\left(U_{n,m+1}-U_{n+1,m}-\sqrt{U_{n+1,m}^2-4U_{n+1,m+1}}-\sqrt{U_{n,m+1}^2-4U_{n,m}}\right)\\+&\chi_{n+m}\left(U_{n,m}-U_{n+1,m+1}-\sqrt{U_{n+1,m+1}^2-4U_{n+1,m}}-\sqrt{U_{n,m}^2-4U_{n,m+1}}\right)=0.\label{td}}
Unlike all the other examples presented in this paper, this equation is not polylinear and even not polynomial. Equations of this kind, with the square roots, have been discussed e.g. in \cite{an13}. 

A generalized symmetry in the $n$-direction, obtained from \eqref{V} by the invertible transformation \eqref{uU}, reads:
\eqs{ &\frac{dU_{n,m}}{dt_1}=(\chi_{n+m}-\chi_{n+m+1}U_{n,m})(U_{n+1,m}-U_{n-1,m}).\label{stn}}
In any fixed point $m$ this equation is equivalent, up to the shift and rescaling $U_{n,m}$, to eq. \eqref{Volt} with $P_n(u_{n})=\chi_{n}u_{n}+\chi_{n+1}.$ So, eq. \eqref{td} defines a chain of B\"acklund  auto-transformations for the Toda lattice equation. 

A symmetry in the $m$-direction obtained from \eqref{fir_V} is of the form:
\eqs{ &\frac{dU_{n,m}}{dt_2}=\frac{C_m\chi_{n+m+1}\left(U_{n,m-1}U_{n,m+1}-4U_{n,m}+\sqrt{U_{n,m-1}^2-4U_{n,m}}\sqrt{U_{n,m+1}^2-4U_{n,m}}\right)}{2(U_{n,m+1}-U_{n,m-1})}\\&+\frac{\chi_{n+m}C_m\left(U_{n,m}^2-4U_{n,m+1}+\sqrt{U_{n,m}^2-4U_{n,m+1}}\sqrt{U_{n,m}^2-4U_{n,m-1}}\right)}{2(U_{n,m+1}-U_{n,m-1})}+\chi_{n+m}C_{m+1},\label{tsymm}}
where $C_m = \alpha m + \beta$. Eq. \eqref{tsymm} with $C_m\equiv m$ provides the master symmetry for eq. \eqref{tsymm} with $C_m\equiv 1$.

If we choose  $C_m\equiv 2$ and introduce $V_m=U_{n,m}$ for any fixed $n$, we are led to the following equation up to the shift of $m$:
\eqs{\frac{dV_{m}}{dt_2}=\frac{R_m(V_{m+1},V_{m},V_{m-1})+\sqrt{R_m(V_{m+1},V_{m},V_{m+1})}\sqrt{R_m(V_{m-1},V_{m},V_{m-1})}}{V_{m+1}-V_{m-1}},\\
R_m(x_1,x_2,x_3)=\chi_{m}(x_2^2-2x_1-2x_3)+\chi_{m+1}(x_1x_3-4x_2).}
This equation is, probably, a new nonautonomous generalization of the well-known integrable lattice equation presented in \cite{y83,y06} .

The formulas \eqref{uU} define two Miura type transformations of the discrete equation \eqref{bacV_1} into \eqref{td} and of the symmetry \eqref{V} into \eqref{stn}. Each of these transformations converts any solution $u_{n,m}$ of \eqref{bacV_1} into a solution of eq. \eqref{td}. These formulas also define transformations of the symmetry \eqref{fir_V} into \eqref{tsymm}, but only on the solutions of the discrete equation \eqref{bacVmTT}. The inverse transformations \eqref{Uu} define two Miura type transformations which relate eqs. \eqref{td} and \eqref{bacVmTT} as well as their symmetries \eqref{tsymm} and \eqref{fir_V}.
The formulas  \eqref{uv} and \eqref{vu} define Miura type transformations of discrete equations and their symmetries as well. 

\section{Conclusions}

In this paper we have constructed a number of new examples of  integrable nonautonomoues discrete equations of the form \eqref{gF}. 

In Section \ref{sec_bac} we describe all polylinear nonautonomous B\"acklund auto-trans\-for\-ma\-tions for the differential-discrete equations (\ref{Volt}-\ref{fP1}) of the Volterra and Toda type. In this way we obtain discrete equations possessing one generalized symmetry of the form \eqref{n-sym}. In Section \ref{sec_sec} we find cases in which there is the second generalized symmetry \eqref{m-sym} and obtain in this way new integrable examples. In Section \ref{sec_tran} we obtain some additional examples, using transformations invertible on the solutions of discrete equations. 

As a result we get discrete equations of the sine-Gordon (\ref{bacV_1},\ref{dw1},\ref{td}) and Liouville (\ref{bacV_2},\ref{bacmV3_1},\ref{bacmV3_2}) type. For Liouville type equations, which are Darboux integrable, we construct the general solutions (\ref{genu},\ref{sol_V3_1},\ref{sol_dw}) by using a special linearization, see for details Section \ref{sec_gen}. The most interesting example of the Liouville type is eq. \eqref{bacmV3_2} which has more complicated first integrals \eqref{fimV3}. Among sine-Gordon type equations, the most interesting from our point of view example is eq. \eqref{dw1}. For this equation we find an $L-A$ pair and construct two  hierarchies of conservation laws and generalized symmetries. The discrete equation \eqref{td} is the only example which is not polylinear. It generates a chain of B\"acklund auto-transformations for the Toda lattice equation. Its second symmetry provides a new nonautonomous generalization of the well-known integrable differential-discrete equation.

In Section \ref{sec_lax} a discrete analog of the well-known dressing chain has been constructed. We find for it an $L-A$ pair and one hierarchy of generalized symmetries and conservation laws. In particular, this example includes eq. \eqref{dw1} and provides it by an $L-A$ pair.

We also have found two degenerate examples of discrete equations, see Sections \ref{Volterracase} and \ref{modi}. One of their symmetries is degenerate in the sense that $\psi_{n,m}$ of \eqref{m-sym} may be zero in some points. Examples of this kind seem to be very close to integrable ones and, in our opinion, deserve further study too.

\ack This work has been supported by the Russian Foundation for Basic Research (grant numbers: 13-01-00070, 14-01-97008-r-povolzhie-a).

\end{document}